\documentclass[conference]{IEEEtran}
\usepackage{amsmath,amssymb,amsfonts}
\usepackage{algorithmic}
\usepackage{array, tabularx}
\usepackage{arydshln}
\usepackage{multirow}
\usepackage{makecell}
\usepackage{dcolumn}
\newcolumntype{d}[1]{D{.}{.}{#1}}
\newcommand\mc[1]{\multicolumn{1}{c}{#1}}

\usepackage[export]{adjustbox}
\usepackage{graphicx}
\usepackage{textcomp}
\usepackage[utf8]{inputenc}
\usepackage{xcolor}
\def\sym#1{\ifmmode^{#1}\else\(^{#1}\)\fi}
\def\BibTeX{{\rm B\kern-.05em{\sc i\kern-.025em b}\kern-.08em
    T\kern-.1667em\lower.7ex\hbox{E}\kern-.125emX}}

\begin{document}

\newenvironment{conditions}
  {\par\vspace{\abovedisplayskip}\noindent
   \begin{tabular}{>{$}l<{$} @{} >{${}}c<{{}$} @{} l}}
  {\end{tabular}\par\vspace{\belowdisplayskip}}

\newenvironment{conditions*}
{\par\vspace{\abovedisplayskip}\noindent
\tabularx{\columnwidth}{>{$}l<{$} @{}>{${}}c<{{}$}@{} >{\raggedright\arraybackslash}X}}
{\endtabularx\par\vspace{\belowdisplayskip}}

\renewcommand{\arraystretch}{1.2}

\title{USeR: A Web-based User Story eReviewer for Assisted Quality Optimizations}

\makeatletter
\newcommand{\linebreakand}{
  \end{@IEEEauthorhalign}
  \hfill\mbox{}\par
  \mbox{}\hfill\begin{@IEEEauthorhalign}
}
\makeatother

\author{
  \IEEEauthorblockN{Daniel Hallmann}
  \IEEEauthorblockA{\textit{University of Bamberg} \\
    \textit{HTW Dresden}\\
    Bamberg, Germany \\
    daniel.hallmann@uni-bamberg.de}
  \and
  \IEEEauthorblockN{Kerstin Jacob}
  \IEEEauthorblockA{\textit{University of Bamberg} \\
    Bamberg, Germany \\
    kerstin.jacob@uni-bamberg.de}
  \and
  \IEEEauthorblockN{Gerald L{\"u}ttgen}
  \IEEEauthorblockA{\textit{University of Bamberg} \\
    Bamberg, Germany \\
    gerald.luettgen@uni-bamberg.de}
  \linebreakand
  \IEEEauthorblockN{Ute Schmid}
  \IEEEauthorblockA{\textit{University of Bamberg} \\
    Bamberg, Germany \\
    ute.schmid@uni-bamberg.de}
  \and
  \IEEEauthorblockN{R{\"u}diger von der Weth}
  \IEEEauthorblockA{\textit{HTW Dresden} \\
    Dresden, Germany \\
    ruediger.von-der-weth@htw-dresden.de}
}

\maketitle

\begin{abstract}
User stories are widely applied for conveying requirements within agile software development teams. Multiple user story quality guidelines exist, but authors like Product Owners in industry projects frequently fail to write high-quality user stories. This situation is exacerbated by the lack of tools for assessing user story quality. In this paper, we propose User Story eReviewer (USeR) a web-based tool that allows authors to determine and optimize user story quality. For developing USeR, we collected 77 potential quality metrics through literature review, practitioner sessions, and research group meetings and refined these to 34 applicable metrics through expert sessions. Finally, we derived algorithms for eight prioritized metrics using a literature review and research group meetings and implemented them with plain code and machine learning techniques. USeR offers a RESTful API and user interface for instant, consistent, and explainable user feedback supporting fast and easy quality optimizations. It has been empirically evaluated with an expert study using 100 user stories and four experts from two real-world agile software projects in the automotive and health sectors.
\end{abstract}


\section{Introduction}

Since introduced by Connextra \cite{Agilealliance:2022us} in 2001, user stories have been widely applied for communicating requirements in agile software projects between authors, such as Product Owners in Scrum \cite{Schwaber:2020th}, and developers. Multiple guidelines (e.g., \cite{Jeffries:2001es, Wake:2003in, Cohn:2004us}), exist to help authors write high-quality user stories, but authors in industry projects frequently fail to comply with these guidelines. Quality issues include, e.g., misleading personas \cite{Wake:2012as}, interdependent user stories \cite{Cohn:2004us}, or skipping parts such as acceptance criteria \cite{Pichler:20135c}.

\begin{table*}
	\centering
	\caption{Real-World Examples of Quality Issues in User Stories from a Health and Automotive Agile Software Project.}
	\label{tab:examples}
	\begin{tabular}{ll}
		\hline
		\textbf{Low-quality User Story\textsuperscript{a}} & \textbf{Issues Description} \\
		\hline
		As a doctor, I need to be able to search for a medicine by its trade name. & Misses the rational \{why\}. \\
		$[...]$ The user wants the prescription software $[...]$. & Includes ``user'' as an unspecific \{persona\}. \\
		As an administrator, I would like to complete the $[...]$, so that $[...]$. & Missing \{acceptance criteria\}. \\
		Company-Bundle Doctrine Upgrade to 2.5. & Focus on technical maintenance tasks. \\
		The list view of translation exports is ported to Ext 5. & No vertical application slice, and uses technical terms. \\
		Drug prescription user data license plate status. $[...]$. & Unspecific terms and unclear meaning. \\
		K4-150 AVWG, MXX FHIR search standardize FHIR search. $[...]$. & Intense use of abbreviations. \\
		$[...]$ clarify technical implementation, questions, and needs PBIs prepared. & Contains questions, to-dos, and prerequisites. \\
		Review of all unit tests. The testers want a coherent scheme for unit tests. & Defines operational and quality development tasks. \\
		$[...]$ Add, remove, activate $[...]$ users from the user group. & Contains multiple functionalities in one user story. \\
		Attribute function group. & Few functionality contexts given with three words in one sentence. \\
		The ASP operator $[...$ Continuing 826 words, $\sim$2 letters$]$ be guaranteed. & Long text used to describe the feature. \\
		Process ePrescription error letter, Error letter in the appendix. & Additional appendix defined, which increased complexity. \\
		\hline
		\multicolumn{2}{l}{\textsuperscript{a} User stories translated from German to English} \\
	\end{tabular}
\end{table*} 

We also show selected user stories from our user story backlogs with quality issues in Table~\ref{tab:examples}. It covers problems with e.g., technical-focused user stories, usage of unspecific terms with unclear meanings and user stories with open questions, to-dos and prerequisites. It is crucial for a project's success that the requirements foster the forming of a shared mental model \cite{Converse:1993wq} within the development team \cite{Sommerville:2005an, Lucassen:2016th}. Low-quality user stories can lead to lengthy discussions, and divergent views between the team members, which increases the risk of incorrect feature implementation and expensive rework.

Based on Femmer and Vogelsang \cite{Femmer:2019re} we understand user story quality by focusing on the stories' purpose in agile software projects. We define user story quality as the degree to which the system stakeholders' needs and the derived consensus are communicated correctly and completely between the customer and the development team and the degree to which the user story enables the development team to execute story-based activities, such as prioritization and estimation.

Support tools for writing user stories (e.g., \cite{Lucassen:2015fo, Lucassen:2016im, Lucassen:2017im}) focus on specific language patterns, such as the conjunction ``and'' as a metric for a non-atomic user story. Hayes et al. \cite{Hayes:2015me} uses machine learning approaches to predict requirements testability. Hallmann \cite{Hallmann:2020id} proposed a research model to evaluate the impact of user story quality on the forming of a shared mental model. In addition, Lai \cite{Lai:2017au} proposes a user story quality measurement model for reducing agile software development risk.

Quality literature \cite{Lucassen:2016im, Lindland:1994un} distinguishes between (i) syntactic quality, which focuses on the textual structure of a user story without considering its meaning and (ii) semantic quality, which addresses the relations and meaning of the user story. Existing tools do not consider semantic criteria of a user story according to Natural Language Processing (NLP) and embedding models \cite{Reimers:2019se, Aaron:2024ge} approaches. Nor do they put different quality metrics in relation, which makes it difficult for authors to receive consistent feedback.

Moreover, machine learning approaches \cite{Hayes:2015me} may fail to provide authors with explainable quality suggestions as the inner workings of Artificial Intelligence (AI) systems become increasingly complex \cite{Lent:2004ae, Tickle:1998th, Edwards:2017sl}. Additional approaches \cite{Hayes:2015me} and existing tools \cite{Lucassen:2016im} capture requirements quality measures and provide their quality results through a report generator, but do not provide a user interface for instant feedback and optimization options.

To support authors to consistently write high-quality user stories that foster the forming of a shared mental model, one needs a comprehensive understanding of user story quality and tools to measure quality. For such support tools to effectively assist the writing of high-quality user stories, it is essential, that the known quality criteria for user stories can be reliably and validly identified with metrics in a tool. This assistance tool can then be used in two ways:

\begin{enumerate}
	\item Providing ratings along the quality criteria for a given user story.
	\item Even pinpoint specific problems and suggest solutions.
\end{enumerate}

\noindent
Inspired by addressing the definition of Femmer and Vogelsang \cite{Femmer:2019re} and the shortcomings of current tools, we propose USeR, a user story eReviewer that allows us to determine user story quality with metrics and offers instant, consistent, and explainable feedback to support authors in writing high-quality user stories. In this paper, we provide four novel main contributions to the existing literature:

\begin{enumerate}
	\item Collecting and refining expert-validated metrics for measuring user story quality.
	\item Details about the theoretical foundations and implementations of the computational metrics.
	\item Technical details of the architecture encompassing the user interface, Application Programming Interface (API), and environment.
	\item An empirical evaluation of USeR in two real-world agile software projects.
\end{enumerate}

\section{Related Work}
\label{section:related_work}

This section introduces the foundations for USeR and summarizes the related work on user stories, user story quality and quality assessment approaches.

\subsection{User Stories}

User stories were introduced by Connextra \cite{Agilealliance:2022us} in 2001 as small text documents to define product features. The ``Connextra Story Card'' contains a \{pattern\} template in the form \{title\}--As \{persona\}, I will \{what\} to \{why\}--\{acceptance criteria\}--\{attachments\}.

The title provides a summary to help the reader elicit a quick feature impression. The persona addresses the user who wants to achieve a specific goal. The what pattern formulates the functionality, and the why part addresses details about the feature's purpose. The acceptance criteria list the feature's scope, especially aspects that must be functional after finalizing the implementation. Finally, attachments provide developers with additional information.

The writing, estimation and implementation of user stories have been extensively studied. Estimation techniques to improve effort predictions are formulated in \cite{Miranda:2009si, Abrahamsson:2011pr, Raith:2013id, Popli:2014co, Popli:2014es}. Stettina et al. \cite{Stettina:2011ne, Stettina:2012do} discuss user stories in the context of agile documentation methods and Sharp et al. \cite{Sharp:2009th} show the usage of user stories in daily routines. Lucassen et al. \cite{Lucassen:2016th} conducted a survey and semi-structured interviews on the use and effectiveness of user stories, which highlights the importance of user story quality for team productivity and the quality of work deliverables.

\subsection{User Story Quality}

Well-known guidelines for authors writing high-quality user stories include CCC (Card, Conversation, Confirmation) \cite{Jeffries:2001es}, INVEST (Independent, Negotiable, Valuable, Estimable, Small, Testable) \cite{Wake:2003in}, and Cohn's guidelines \cite{Cohn:2004us}.

According to these guidelines, authors should keep user stories concise, focusing on short requirements descriptions and a feature scope of, at most, a week of work for a few developers. User stories should contain acceptance criteria to prevent infinite development loops and allow feature testing. High-quality stories also inspire conversation, in which developers can negotiate implementation approaches and estimate implementation effort.

The independent quality aspect emphasizes decreasing complexity by separating functionality. Lastly, the value quality aspect reflects the agile method's focus on customer value delivery with each feature \cite{Beck:2001ma}. Authors should empathize with the user goals to maximize the application's usefulness and acceptance.
The guidelines cited above have been derived from practical experience in working with user stories. Lucassen et al. \cite{Lucassen:2016th} found that practitioners consider them helpful for increasing productivity within the team.

\subsection{Measuring User Story Quality}

However, the guidelines' aspects are highly qualitative and not easy to measure. Lucassen et al. \cite{Lucassen:2015fo, Lucassen:2016im, Lucassen:2017im} proposed a framework for user story quality consisting of 13 criteria for syntactic, semantic and pragmatic quality dimensions of user stories, and developed a tool for computationally measuring a subset of the quality criteria. For the latter, the authors focus on specific language patterns, e.g., conjunction ``and'' as an indicator for a non-atomic user story. Their tool provides valuable support for authors of user stories but does not put different quality indicators in relation.

Hayes et al. \cite{Hayes:2015me} researched measuring requirement quality to predict testability. They used machine learning and statistical analysis approaches. They defined requirement measures and used them to empirically investigate the relevant relationship between each measure and requirements testability. Furthermore, they found that the measures are useful to describe testable and non-testable requirements. In addition, their learned model can be used to predict testability of requirements for other systems.

Hallmann \cite{Hallmann:2020id} proposed a research model to evaluate the role of user story quality in forming a shared mental model based on structural equation modeling. It contains a measurement model to predict quality scores with four indicators: formal, lexical, semantic, and saturation quality. This model allows humans to assess the indicator's weights and loadings to understand and optimize the model's behavior.

Lai \cite{Lai:2017au} presented a user story quality measurement model for reducing agile software development risk. The quality factors are split into three concepts which are basic quality, management quality, and acceptance quality. The metrics will be collected individually and form later in a linear combination the overall quality score.

\section{USeR - A User Story eReviewer}
\label{section:USeR}

In this section, we provide a tool to support authors in the handling of user stories. Especially we will give support to e.g., Product Owners in writing high-quality user stories. Our motivation is driven by two primary objectives:

\begin{itemize}[\IEEEsetlabelwidth{O1.}]
	\item[O1.] USeR should support a full computational measurement of user story quality criteria.
	\item[O2.] USeR should provide a user interface for instant, consistent, and explainable user feedback.
\end{itemize}

\begin{figure}[htbp]
	\centerline{\includegraphics[width=0.5\textwidth]{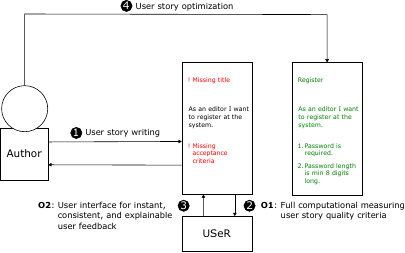}}
	\caption{Conceptual framework of USeR. The system facilitates an iterative process where authors receive automated quality feedback on user stories and subsequently implement optimizations to improve their quality.}
	\label{fig:concept}
\end{figure}

\noindent
We outline the concept of USeR in Figure~\ref{fig:concept}. The author writes first an initial version of the user story, USeR secondly will measure the user story quality based on different quality criteria. Third, USeR provides metrics as instant feedback of the user story quality. In the fourth step, the author can decide to make changes to the user story. Steps one to four can be iteratively looped until the author decides to accept the user story as final.

\subsection{Metrics Development}

The following section describes the development of the metrics for USeR. We will first describe the process of the metrics development and then the metrics themselves.

\subsubsection{Collect Metrics}

To find reasonable metrics for user story quality we conducted a literature review, practitioners brainstormings and interviews, and research group meetings.

The literature review started on general quality criteria for documents and was then narrowed to requirements documents, and further to requirements in the agile development processes, and guidelines specific to user stories. We identified 14 papers as relevant for user story quality metrics and extracted 27 metrics from the papers.

To find additional suitable measurements for the metrics, we conducted two one-hour brainstorming sessions. Initially, a session was conducted with one Scrum Master, one Product Owner, two senior developers, and one developer. Subsequently, a second session was undertaken, comprising eight senior developers, one Product Owner, and one Scrum Master. We identified 19 metrics during the brainstorming sessions.

Finally, during research meetings, we developed a set of metrics in our research group. As a starting point, we used the previous literature review and brainstorming results to guide the metrics development. We collected 31 metrics from the research group sessions. In total, we gathered 77 potential quality metrics.

\subsubsection{Select Metrics}

\begin{table*}
	\centering
	\caption{Summary of the Collected and Prioritized Quality Metrics for USeR.}
	\label{tab:metrics_overview}
	\begin{tabular}{lllll}	
		\hline
        \textbf{Metric} & \textbf{Description} & \textbf{Source}\textsuperscript{a} & \textbf{References} & \textbf{USeR\textsuperscript{b}} \\
		\hline
		\textbf{Readable related} & & & & \\
		Readable & Measuring the readability. & L & \cite{DeSouza:2005as,Flesch:1948an,Stettina:2013th} & Y \\
		Easy Language & Measuring the language complexity. & E & & Y \\
		Word Sparse & Measuring the orientation to an average word count. & R & & Y \\
		Sentence Sparse & Measuring the orientation to an average sentence count. & R & & Y \\
		Language Differences & Evaluating language differences (e.g., English\textsuperscript{c}). & E & & \\
		Stop Word Story & Evaluating the number of stop words (e.g., and, or). & R & & \\
		Conjunction Story & Evaluating the number of conjunctions (e.g., but, either). & R & & \\
		Unspecific Story & Evaluating the number of unspecific terms (e.g., many). & R & & \\
		\textbf{Customer related} & & & & \\
		Customer Speak & Checking the story is written in the customer language. & R & & Y \\
		Valuable & Evaluating the valuable criteria of a user story. & L & \cite{Cohn:2004us,Wake:2003in,Abrahamsson:2011pr} & \\
		\textbf{Format related} & & & & \\
		Format Complete & Evaluating the presence of filled-in patterns. & R & & Y \\
		Correct Persona & Checking whether the persona is a valid persona. & E & & \\
		Has Additionals & Checking additional information beside default patterns. & E & & \\
		Has Correct Order & Checking the pattern order (e.g., 1. Title, 2. Persona). & E & & \\
		Uniform, Remain Format & Checking the format is consistent across user stories. & L & \cite{Lucassen:2016im} & \\
		\textbf{Understandable related} & & & & \\
		Complete & Measuring the completeness (e.g., no TBD\textsuperscript{d} inside). & E &  \cite{Boehm:1984ve, Davis:1993sr,DeSouza:2005as, Lucassen:2016im, Iso:2018is} & \\
		Understanding & Measuring the understanding of the user story. & E & & \\
		\textbf{Complexity related} & & & & \\
		Independent, Atomic & Evaluating the independent criteria of a user story. & L & \cite{Cohn:2004us, Davis:1993sr,Iso:2018is,Lucassen:2015fo,Wake:2003in} & Y \\
		Small & Evaluating the small criteria of a user story. & L & \cite{Wake:2003in} & Y \\
		Consistent, Conflict-free & Evaluating the consistent criteria of a user story. & L & \cite{Iso:2018is,Boehm:1984ve,Davis:1993sr,Flesch:1948an,Lucassen:2015fo} & \\
		Story Points & Measuring the story points (e.g., 1. Feasible, 20. Too big). & R & & \\
		Traceable & Evaluating the traceable criteria of a user story. & L & \cite{Davis:1993sr, Iso:2018is, Malone:1994th} & \\
		\textbf{Testable related} & & & & \\
		Acceptance Criteria Quality & Evaluating the quality of acceptance criteria. & E & & \\
		Testable & Measuring the number of acceptance criteria. & L & \cite{Wake:2003in} & \\
		\textbf{Estimable related} & & & & \\
		Measurable & Evaluating the measurable criteria of a user story. & L & \cite{Davis:1993sr,Iso:2018is,Jeffries:2001es, Malone:1994th, Wake:2003in} & \\
		Feasible, Attainable, Realizable & Evaluating the feasible criteria of a user story. & L & \cite{Davis:1993sr,Iso:2018is, Malone:1994th} & \\
		Small Spread & Measuring the spread of the 3-point estimation. & E & & \\
		\textbf{Interaction related} & & & & \\
		Views & Measuring clicks, views, comments in e.g., Jira. \cite{Atlassian:2002ji}. & E & & \\
		\textbf{Up-to-date related} & & & & \\
		Up-to-date & Evaluating up-to-date criteria of a user story. & L & \cite{DeSouza:2005as} & \\
		Volatile & Evaluating the volatile criteria of a user story. & L & \cite{Davis:1993sr} & \\
		Time Current Update & Measuring the last update and current span. & R & & \\
		Time Estimation Update & Measuring the last update and estimation span. & R & & \\
		Time Development Update & Measuring the last update and development start span. & R & & \\
		Time Development Estimation & Measuring the last estimation and development start span. & E & & \\
		\hline
		\multicolumn{4}{l}{\textsuperscript{a} Source (Literature Review (L), Expert Brainstorming's/Interviews (E), Research Group Meetings (R)), \textsuperscript{b} Implemented in USeR (Y = Yes),}\\
		\multicolumn{4}{l}{\textsuperscript{c} American and British English, \textsuperscript{d} To Be Defined}\\
	\end{tabular}
\end{table*}

To choose the most relevant metrics as potential candidates, we conducted two expert Delphi sessions. The first session spans two hours with one manager, three Scrum Masters, and seven developers. The second fine-tuning session covers a one-and-a-half slot with four developers. In total, we selected 34 metrics and started implementing eight metrics from the top prioritized groups. Table~\ref{tab:metrics_overview} summarizes the prioritized and first implemented quality metrics.

\subsection{Metrics}

After the theoretical formation of the metrics, we started with deriving algorithms, which allow measuring values that can be directly used to interpret the quality. The following section describes the metrics with their equations and the rationale behind them.

\subsubsection{Format Complete}

The format completeness metric focuses on a complete definition, in the sense, that all patterns of the user story from the Connextra template \cite{Agilealliance:2022us} are filled in to ensure a minimum of information to understand the user story. The presence of the patterns is essential to capture the sense of the user story, start a discussion, and find solutions and estimates regarding the implementation and necessary effort.

\begin{equation}
	Format\ Complete = 
	\frac{1}{n} 
	\sum_{i=1}^{n}{
		\begin{cases}
			1, \text{if}\ pattern_{i} \not = \text{""} \\
			0, \text{else} \\
	\end{cases}} 
	\label{eq:format_complete}
\end{equation}
where:
\begin{conditions}
	pattern_{i} & \in & \{title, persona, what, why, acs, attachments\} \\
	n & = & total number of patterns
\end{conditions}

We calculate $Format\ Complete$ in \eqref{eq:format_complete} by identifying the total number of filled-in patterns an author has defined in a user story. Subsequently, we build the fraction between the number of filled-in patterns and the total number of patterns in the Connextra template to determine the completeness of the user story.

\subsubsection{Readable}

Readability ensures, that developers can decode and extract information to support the comprehension of the user story goals. We see the readability of the text structure related to linguistic and syntactic dimensions as part of user story quality. We measure reading ease based on an approach from Flesch \cite{Flesch:1948an}, who proposed an equation to quantify readability based on text properties such as sentences, words, and syllables.

\begin{equation}
	Readable = \frac{206.835 - 84.6\cdot asw - 1.015\cdot asl}{mf}
	\label{eq:readable}
\end{equation}
where:
\begin{conditions}
	asw & = & average number of syllables per word \\
	asl & = & average sentence length \\
	mf & = & maximal Flesch readable index \\
\end{conditions}

We integrated Flesch's approach in the $Readable$ metric \eqref{eq:readable} unchanged, as it is optimized for calculating readability. The factors ($asw$, $asl$) are calculated based on the individual text, as opposed to the constants and $mf$, that were calibrated on texts during implementation and calculated on an empty string as the simplest readable sign. We selected the German version because of our German user story backlogs.

\subsubsection{Customer Speak}

We see the domain language as part of user story quality in addressing a textual user story quality dimension. This metric describes the domain focus that a user story represents. We focused on the story author's usage of the domain language to ensure that the user story is understandable and relevant to the business domain.

\begin{equation}
	Customer\ Speak = \frac{gw \cap uw}{uw}
	\label{eq:customer_speak}
\end{equation}
where:
\begin{conditions}
	gw & = & total glossary words \\
	uw & = & unique user story words
\end{conditions}

In the $Customer\ Speak$ equation \eqref{eq:customer_speak}, we first generated a glossary through an automatically evaluating of all backlog user stories to identify customer business domain keywords. Special words are interesting because they have dedicated meanings in the business domain. Next, all domain keywords were detected in the user story under test. Finally, we calculate the ratio of the domain keywords found and the unique number of user story words.

\subsubsection{Small}

User stories should be small to ensure an easy understanding and implementation \cite{Wake:2003in}. A user story should contain a low number of dependent topics to make the development less complex with fewer code changes on fewer locations. Focus on one topic can also reduce the mental load for developers to better identify unknowns and failure risks which can prevent breaking another application source.

\begin{equation}
	Small = 1 - \frac{1}{n} \sum_{i=1}^{n}{
		\begin{cases}
			1, \text{if}\ prob_{i} \geq thr \\
			0, \text{else} \\
		\end{cases}}
	\label{eq:small}
\end{equation}
where:
\begin{conditions}
	n & = & total number of topics in the backlog \\
	prob & = & probabilities of topics occurring in a user story \\
	thr & = & threshold to associate a topic with a user story \\
\end{conditions}

We calculate the $Small$ metric \eqref{eq:small} by identifying the probabilities of topics occurring in a user story. We use a topic model to identify the topics in the user story, and we estimate the probability for each topic, whether it is present in the user story. To associate a topic with a user story, a threshold is used. Finally, we calculate the fraction between the number of topics found in the user story and the total number of backlog topics.

\subsubsection{Independent}

Wake's \cite{Wake:2003in} approach proposed that user stories should be separated by functionality. It could be more complex to propose development approaches and make estimations for user stories that are not independent, because of dependencies on existing features and code, which can lead to higher effort to integrate the new user story.

\begin{equation}
	Independent = 1 - \frac{1}{n} \sum_{i=1}^{n} sim(B_{i}, us)
	\label{eq:independent}
\end{equation}
where:
\begin{conditions}
	n & = & total number of user stories in the backlog \\
	B & = & backlog of user stories \\
	us & = & user story under test \\
	sim & = & cosine similarity between user story $B_i$ and $us$
\end{conditions}

We measure $Independent$ in equation \eqref{eq:independent} by the cosine similarity between the user story under test and all user stories in the backlog. In addition, we calculate the ratio between the number of user stories in the backlog and the sum of the cosine similarity between the user story under test and all backlog user stories.

\subsubsection{Word/Sentence Sparse}

We focus on the sparse metrics on the average user story length to provide the best fitting text length for developers in the project. Limit text can prolong the discussion time until the first ideas emerge. Furthermore, it is recommended to avoid including excessive text to keep the user story open for discussion and collaboration \cite{Jeffries:2001es, Cohn:2004us}.

\begin{equation}
	Word/Sentence\ Sparse  = \frac{n - w}{m - w}
	\label{eq:sparse}
\end{equation}
where:
\begin{conditions}
	n & = & n words, sentences of the user story under test \\
	w & = & n min or max words, sentences found in the backlog \\
	m & = & n mean words, sentences found in the backlog \\
\end{conditions}

We calculate $Word/Sentence\ Sparse$ \eqref{eq:sparse} by calculating the difference between the number of words or sentences and the minimum or maximum number of words or sentences in the backlog. In addition, we build the difference between the mean number of words or sentences and the minimum or maximum number of words or sentences in the backlog. Finally, we determine the fraction between both differences to find the ratio of how far or close the user story has an orientation to the average number of words or sentences in the backlog.

\subsubsection{Easy Language}

The metric covers the usage of basic common words in the user story with a semantic focus. Common words (e.g., doctor, machine) used to write the content of the user story can help to understand the meaning of the words and the aims of the user story faster. Common words are learned early, used frequently, and bounded on many examples, feelings and emotions which represent a solid presentation of the words' purpose \cite{Langacker:1987fo}.

\begin{equation}
	Easy\ Language = \frac{tw \cap uw}{uw}
	\label{eq:easy_language}
\end{equation}
where:
\begin{conditions}
	tw & = & total basic words \\
	uw & = & unique user story words
\end{conditions}

We measure $Easy\ Language$ \eqref{eq:easy_language} based on a list of common German words \cite{ZSL:2023gr}. First, we calculate the list of unique words included in the user story. Next, we build the intersection of unique words and common words to find the overlapping in the user story. Finally, we calculate the fraction between the number of common words and unique words in the user story to identify the ratio of easy words used against other more complex words.

To ensure consistent behavior of USeR to allow the authors to make consistent changes to the user story to improve the quality, we set the scale type of all the metrics as an interval scale ranging from 0\% to 100\%. 0\% represents a low-quality user story and 100\% represents a high-quality user story.

\subsection{Architecture}

The architecture of USeR is divided into three main components: (i) web app, (ii) API, and (iii) environment.

\begin{figure}[htbp]
    \centerline{\includegraphics[width=0.4\textwidth]{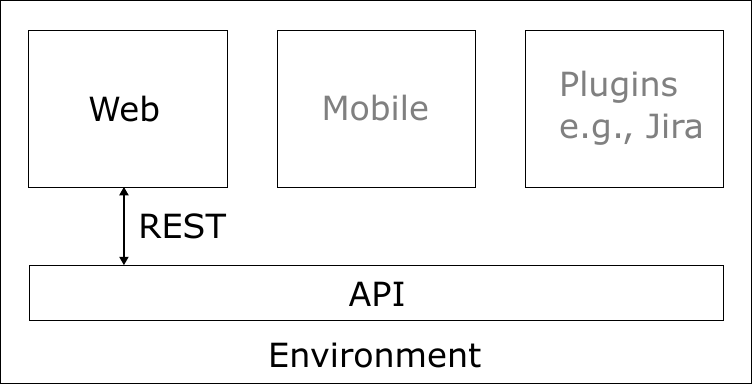}}
    \caption{Architecture of USeR containing the web-based user interface and the API as backend. Virtualization is used to host the user interface and the API. USeR implemented a Representational State Transfer (RESTful) API to facilitate standardized communication with the web app.}
    \label{fig:architecture}
\end{figure}

\begin{table}
    \centering
    \caption{USeR Integrated Tools, Tool Version, and Sources.}
    \label{tab:architecture}
    \begin{tabular}{llll}
        \hline
        & \textbf{Tool} & \textbf{Version} & \textbf{Source} \\
        \hline
        \textbf{Web} & Node.js & 20.6.1 & https://nodejs.org \cite{NodeJS:2024no} \\
        & React & 18.2.0 & https://react.dev \cite{React:2024no} \\ 
        \hline
        \textbf{API} & Python & 3.10.13 & https://python.org \cite{Python:2024no} \\
        & spaCy & 3.6.1 & https://spacy.io \cite{SpaCy:2024sp} \\
        & sklearn & 1.3.0 & https://scikit-learn.org \cite{scikit-learn:2011sc} \\
        & BERTopic & 0.15.0 & BERTopic\textsuperscript{a} \cite{Grootendorst:2022be} \\
        \hline
        \textbf{Environment} & Docker & 24.0.6 & https://docker.com \cite{Docker:2024no} \\
        \hline
		\multicolumn{3}{l}{\textsuperscript{a} https://github.com/MaartenGr/BERTopic}
    \end{tabular}
\end{table}

Figure~\ref{fig:architecture} shows the architecture of USeR. The web-based user interface allows the authors to interact with the tool. The API is the back-end part, which includes processing the user stories and calculating the quality metrics. The environment is the infrastructure part, which hosts the user interface and the API. We present an overview of the tools and tool versions applied in USeR in Table~\ref{tab:architecture}.

\subsection{API}
\label{section:api}

The API is composed of four distinct components: (i) data importer, (ii) training, (iii) prediction, and (iv) interpretation.

\begin{figure*}[htbp]
    \centerline{\includegraphics[width=1.0\textwidth]{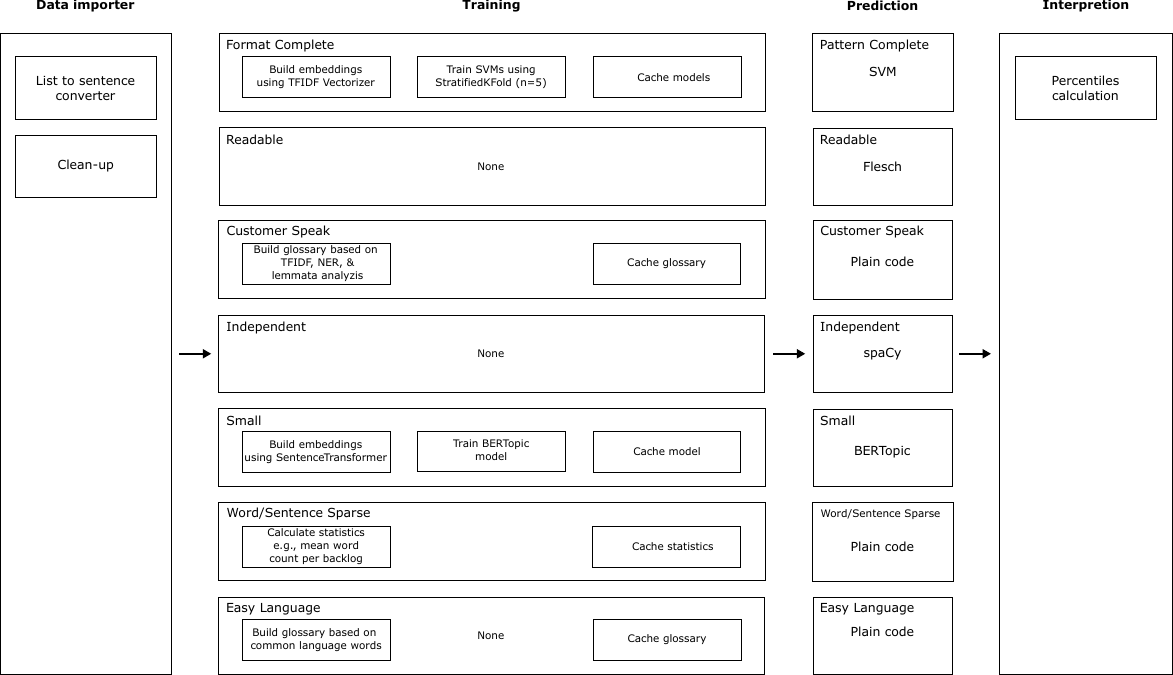}}
    \caption{USeR's API parts: data importer that covers text cleaning and converting, training models and metrics preparation, prediction of the metrics values, and interpretation with calculating the percentiles as orientation for authors to plan their user story quality optimizations.}
    \label{fig:api}
\end{figure*}

\subsubsection{Step 1: Data importer}

The data importer in Figure~\ref{fig:api} covers two main functions: text parsing and cleaning up. We extract the stories from our management tools (e.g., Jira \cite{Atlassian:2002ji}) which provide an API to retrieve Comma Separated Values (CSV) files. In addition, we clean special characters and reformat the text.

\subsubsection{Step: 2: Training}

The second step covers the training of the language models and the preparation of glossaries and text statistics for the calculation of the quality metrics. First, we prepare the text embeddings with a Term Frequency-Inverse Document Frequency (TF-IDF) vectorizer \cite{scikit-learn:2011sc}, which we use to train a Support Vector Machine (SVM) for classification tasks in the \textit{Format Complete} metric and a Bidirectional Encoder Representations (BERTopic) model for identifying topics in user stories in the \textit{Small} metric.

In addition, we build the \textit{Customer Speak} glossaries by selecting words in three different parsing algorithms based on NLP technologies: TF-IDF embeddings, Named Entity Recognition (NER), and lemmata analysis. We build the \textit{Easy Language} glossaries by integrating a common German basic word list \cite{ZSL:2023gr}.

Finally, we calculate with spaCy the minimum, mean, and maximum word and sentence count for each project user story in our backlogs to prepare the prerequisite values for measuring the \textit{Word/Sentence Sparse} metrics. To optimize performance, we cache all models, glossaries, and statistics.

\subsubsection{Step: 3: Prediction}

We use a two-step sequence to predict the final quality metrics: (1) determine the raw values used as variables in the metrics equations through predictions of the trained models (SVM, BERTopic) or plain code extraction (spaCy text similarity functionality, Python Flesch library functions). Further, we use the raw values to (2) calculate the final quality values based on the metrics equations. 

\subsubsection{Step: 4: Interpretation}

We employ a percentile calculation to extract the lower, middle, and upper quartiles of the qualities of all project user stories. This provides ranges for the quality metrics to the authors, thereby supporting the interpretation of the quality predictions of the user stories.

\subsection{User Interface}
\label{section:user_interface}
\begin{figure}[htbp]
	\centerline{\includegraphics[width=0.5\textwidth, frame]{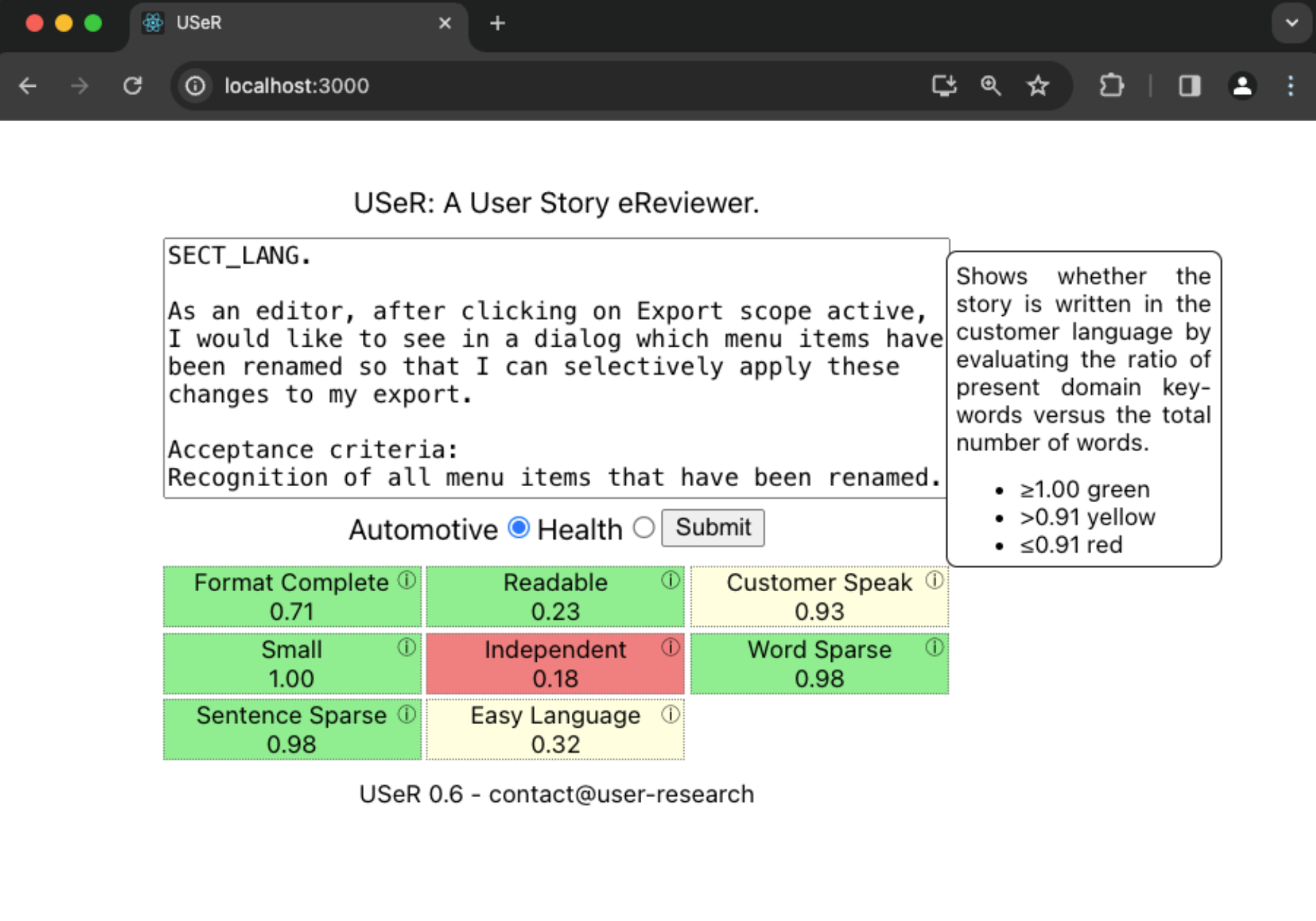}}
	\caption{Web-based user interface of USeR, which is structured into the user story input area and the quality metrics part. Authors can edit user stories in the text box and the quality metrics show the quality ratings of a user story.}
	\label{fig:user_interface}
\end{figure}

The web-based frontend in Figure~\ref{fig:user_interface} is divided into two main components: (i) the user story input area and (ii) the quality metrics part showing the quality ratings of a user story. The user story input field provides an editor for the author to add and change a user story. The quality metrics grid displays each quality metric with its name and rating result. The quality metric is colored based on the percentiles and provides a tooltip with a metric's description and the ranges of the percentiles to guide authors about the degree of changes to optimize the quality level.

USeR's API components in Section~\ref{section:api} are interconnected in a pipeline configuration that runs autonomously by providing the user stories as training data, which allows a full computational measurement of user story quality criteria, supporting our research objective O1. In addition, the user interface in Section~\ref{section:user_interface} allows authors to interact with USeR for instant, consistent, and explainable user feedback, which covers the objective O2.

\section{Evaluation}
\label{section:methodology}

The following section describes our empirical study of USeR's evaluation in two real-world projects. When evaluating a tool, it is common to use experts \cite{Arora:2016ex, Israel:2008in}. To properly set up an expert study, we first performed a literature review on expert studies in computer science. To focus and coordinate our work, we scoped our empirical evaluation of USeR on two research questions:

\begin{itemize}[\IEEEsetlabelwidth{RQ1.}]
	\item [RQ1.] How is USeR's prediction quality?
	\item [RQ2.] What is the best metrics composition?
\end{itemize}

\noindent
First, we aim to assess how accurately USeR can predict the quality ratings of user stories compared to expert ratings. Second, we address finding the optimal combination of metrics for predicting user story quality.

\subsection{Projects and Companies}

Our first project is from a German automotive company, one of the leading international manufacturers of commercial vehicles and transport solutions, with 39,000 employees and 120,000 vehicles sold annually. The development project considered here is for a company-wide documentation solution, which ran between March 2013 and December 2015 with a Scrum team of nine developers, one Product Owner, and one Scrum Master. 

Our second project is from a product company in the German health sector. It is one of the leading companies for Information Technology (IT) solutions for medical practice management, supporting administrative and operational workflows. Since 2019, the project has been developing a new module for medicine prescription with a Scrum team consisting of five developers, one Product Owner, and two quality assurance specialists. The project does not have a Scrum Master.

\subsection{Experts}

We defined a target group of experts who are familiar with user stories. Our focus was on experts who work as developers, quality assurance specialists, or Product Owners in agile software development and handle user stories. We used our contacts to recruit experts from the aforementioned two agile projects.

Based on related work on expert studies \cite{Arora:2016ex, Israel:2008in}, we chose four experts for our study and recruited one developer, one Product Owner, and two quality assurance engineers. We asked the experts about their experiences concerning agile software development, user stories, and their business domain.

\begin{table}
    \centering
    \caption{Experience in years of the Automotive and Health Experts.}
    \label{tab:experts}
    \begin{tabular}{llll}
        \hline
        & \textbf{Agile} & \textbf{User Stories} & \textbf{Domain} \\
        \hline
        E1 & 12 & 12 & 10 \\
        E2 & 14 & 12 & 8 \\
        E3 & 3 & 3 & 3 \\
        E4 & 5 & 8 & 5 \\
        \hline
    \end{tabular}
\end{table}

Table~\ref{tab:experts} presents the years of experience of the two automotive (E1, E2) and two health (E3, E4) experts. The average experience in agile software development and user stories is 9 years. The average experience in the business domain is 7 years.

\subsection{User Stories and Sampling}

We had access to the user story repositories of the two projects. The automotive project had 815 user stories, and the health project had 150 user stories.

We executed a power analysis first to determine the number of user stories necessary to evaluate USeR. Based on our pretests and recommendations \cite{Cohen:1988st, Gpower:2009st} for our planned regression analysis \cite{Casella:2024st}, we determine 50 user stories for each automotive and health project. We selected user stories by randomized sampling from the full project's user story backlogs to ensure representatives across all user stories. Finally, we assigned the automotive and health experts to their automotive and health sets.

\subsection{Questionnaire}

We used a questionnaire \cite{Oates:2006re, Creswell:2017re} to collect the user story ratings from our experts. The questionnaire includes introductions, participants' demographic questions, a tutorial, and the primary quality ratings. The primary quality rating section included 50 user stories with a single rating question:

\begin{enumerate}
	\item How is the quality of the user story?
\end{enumerate}

We chose a broader question with no restrictions to allow experts to express an unbiased rating on the user story quality. We used a unipolar positively polarized 5-point-Likert scale \cite{Likert:1932at} that ranges from one: very low quality to five: very high quality. The questionnaire was realized with LimeSurvey \cite{Limesurvey:2022li}.

Prior to implementation, the questionnaire underwent two sequential validation phases: an initial pretest with six domain experts, followed by a secondary pretest with two additional domain experts.

\subsection{Procedure}

Using the questionnaire, our experts rated the user story quality in rating sessions. We selected guided sessions over a self-administered questionnaire to increase the response rate and to ensure the quality of our expert rating data. The rating sessions took place between December 2023 and February 2024. We started the approximately one-hour sessions with a five-minute briefing, afterward, we guided the experts through the rating phase and closed the sessions with a five-minute debriefing. For the sessions, we created a conducive environment (e.g., enough time and silence). We used Zoom \cite{Zoom:2022zo} as a video and audio tool, and recorded all sessions. Subsequently, we also rated the 50 automotive and 50 health user stories with USeR.

\subsection{Analysis}

To answer both research questions: RQ1 and RQ2, we used multiple linear regression analysis \cite{Myrtveit:1999ac, Abrahamsson:2011pr}. We calculated the means of the two expert ratings, removed outliers and evaluated the inter-rater reliability of the expert \cite{Gisev:2013in}. We used SmartPLS \cite{SmartPLS:2024no} with the parameters; test type: two-tailed, significance level: $\alpha = 0.05$, standard error type: normal standard errors, and weighting vector: none.

The source code for USeR and the results of the expert study are available at the following URL: https://github.com/user-research/USeR.

\section{Results}
\label{section:results}

Our evaluation identified no missing expert ratings, giving us a response rate of 100\%. We attribute the high response rate to our support of the experts during the rating session.

Furthermore, we eliminated outliers with the Interquartile Range (IQR) method \cite{Barnett:1994ou} by setting lower and upper bound limits at a factor of 1.5 \cite{Hoaglin:2003st}. In total, 15 outliers were eliminated from the automotive dataset, and 12 were removed from the health dataset. The presence of multiple outliers generated by the ratings of a single user story led to 42 automotive and 39 health user stories.

\begin{figure}[htbp]
	\centerline{\includegraphics[width=0.3\textwidth]{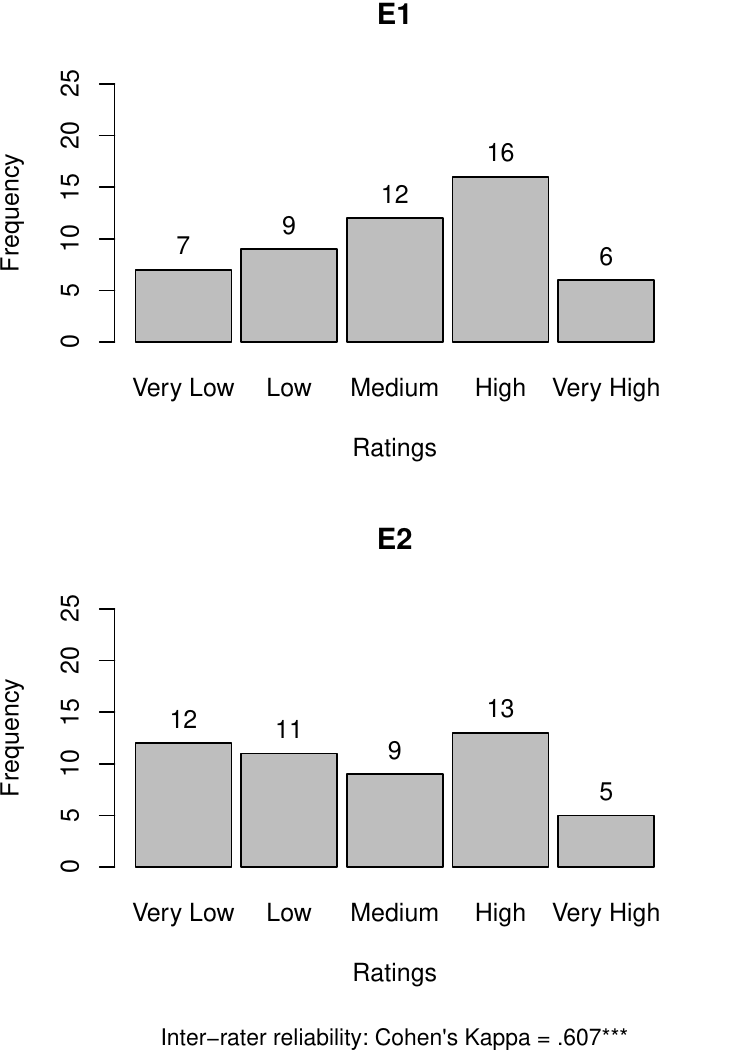}}
	\caption{Automotive experts' statistics show an increased medium-quality rating tendency and a strong inter-rater reliability.}
	\label{fig:automotive_experts_barplot}
\end{figure}

\begin{figure}[htbp]
	\centerline{\includegraphics[width=0.3\textwidth]{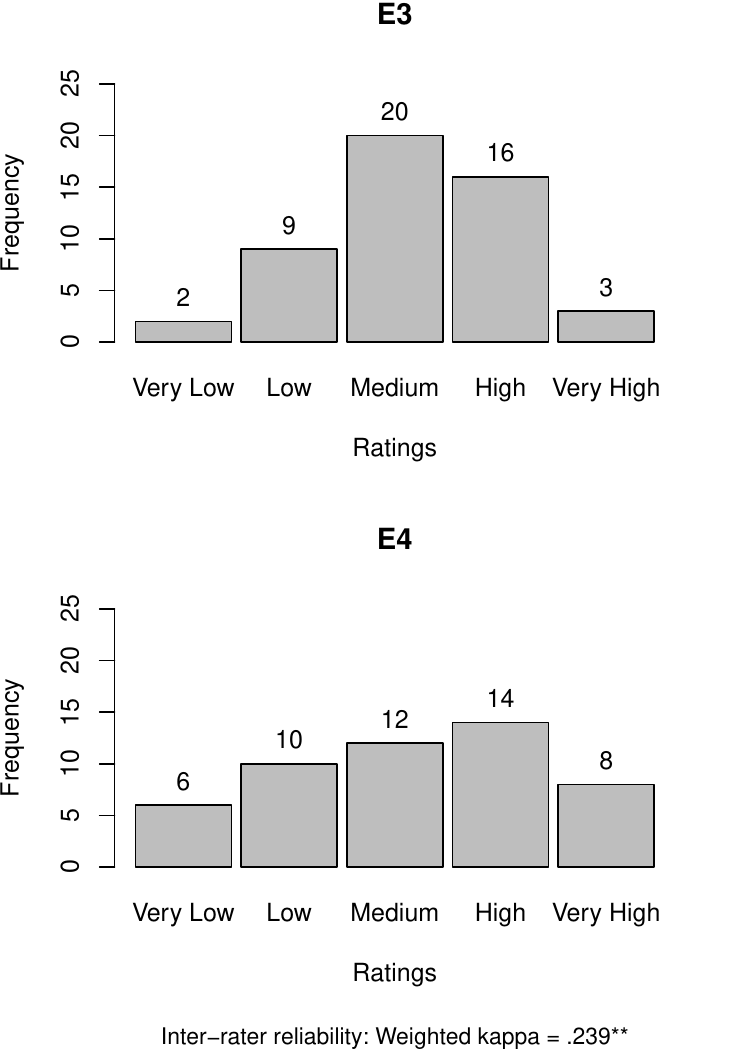}}
	\caption{Health experts' statistics show acceptable inter-rater reliability through the Weighted kappa metrics and agreement in the bar plots. In addition, the experts show an increased medium-quality rating tendency.}
	\label{fig:health_experts_barplot}
\end{figure}

As shown in Figure~\ref{fig:automotive_experts_barplot} and Figure~\ref{fig:health_experts_barplot} and the bar plots, the four experts tend to a slightly increased medium-quality rating tendency. In addition, we found a strong inter-rater agreement for the automotive experts with 61\% and an acceptable inter-rater agreement for the health experts with 24\% at a significance level: $\sym{***}p<.001,\ \sym{**}p<.01,\ \sym{*}p<.05$.

\begin{table*}
	\centering
	\caption{Determination Coefficient, Standardized Beta Weights, and Variance Inflation Factor of the Regression Analysis}
	\label{tab:regression_results}
	\begin{tabular}{l*{9}{d{3.3}}}
		\hline
		& \mc{R\textsuperscript{2a}} & \mc{Format Complete\textsuperscript{b}} & \mc{Readable} & \mc{Customer Speak} & \mc{Small} & \mc{Independent} & \mc{Word Sparse} & \mc{Sentence Sparse} & \mc{Easy Language} \\
		\hline
		A\textsuperscript{c} & 0.76 & 0.35\sym{**} & -0.07 & 0.00 & -0.21\sym{*} & 0.11 & 0.74\sym{**\diamond} & -0.10\sym{\diamond} & 0.01 \\
		H\textsuperscript{d} & 0.44 & 0.56\sym{**} & -0.16 & -0.01 & -0.24 & 0.06 & 0.26 & -0.15\sym{\diamond} & 0.21 \\
		G\textsuperscript{e} & 0.48 & 0.43\sym{***} & -0.18 & 0.01 & -0.19 & 0.14 & 0.52\sym{**\diamond} & -0.11\sym{\diamond} & 0.05 \\
		\hline
		\multicolumn{10}{l}{\textsuperscript{a} Determination coefficient, \textsuperscript{b} Standardized beta weights, \textsuperscript{c} Automotive project, \textsuperscript{d} Health project, \textsuperscript{e} Global (Automotive + Health) project,} \\
		\multicolumn{10}{l}{Significance level: $\sym{***}p<.001,\ \sym{**}p<.01,\ \sym{*}p<.05$, Multicollinearity correlation: $\sym{\diamond}$VIF $> 4.0$} \\
	\end{tabular}
\end{table*}

Table~\ref{tab:regression_results} shows determination coefficients and strong effects of $0.76$ for the automotive project and $0.44$ for the health project. In addition, we ran a multiple regression analysis on a global dataset by combining the automotive and health data. We were motivated to retrieve further insights into how USeR behaves concerning a global perspective. The determination coefficient for the combined analysis is $0.48$, which also results in a strong effect. In addition, we found significant positive standardized beta weights for the metrics \textit{Format Complete}, \textit{Word Sparse} and significant negative weights for \textit{Small}.

We detected an increased multicollinearity correlation with a Variance Inflation Factor (VIF) $> 4.0$ \cite{Fox:2018an} between the metrics \textit{Word Sparse} and \textit{Sentence Sparse}, which we expected because of their close theoretical conception. We prefer to keep the positive significant \textit{Word Sparse} metric and eliminate further the insignificant negative \textit{Sentence Sparse} metric.

\section{Discussion}
\label{section:discussions}

The evaluation, confirming USeR's prediction quality, shows that it accounts for 76\% of the automotive experts' ratings, 44\% of the health experts' ratings, and 48\% of the combined ratings from both projects. Thus, under RQ1, we can deduce that USeR's quality predictions effectively represent expert predictions. This has practical implications, suggesting that USeR can readily assist authors in creating user stories at an expert level.

RQ2 aims at finding the best metrics composition in identifying significant metrics for measuring user story quality. \textit{Format Complete} is the most favorite metric, indicated by the significant results across both projects and on a global level. \textit{Word Sparse} emerges as a significant metric, suggested by high weights in the automotive project and globally. We assume the high significance of both metrics can be explained by the effective computationally detection of syntactical criteria, which aligns with previous results \cite{Lucassen:2015fo, Lucassen:2016im, Lucassen:2017im}.

Insights from the rating session revealed experts align their quality ratings with the user story length and the presence of the filled-in patterns. Lengthy user stories tend to receive low-quality ratings, as do significantly shorter ones. User stories of average length are often rated as high quality. In addition, the presence of filled-in patterns in a user story is often associated with high-quality ratings and the absence of filled-in patterns with low-quality ratings. We assume experts covered first the length of a user story and second the presence of all filled-in patterns. If both quality criteria are present, it is more likely to be rated as a high-quality user story. If one of the quality criteria is missing a low-quality user story rating is more likely.

Our evaluation reveals \textit{Small} as a negative significant metric, with opposed behavior regarding the expert's ratings. Small-rated user stories described a single topic which makes them more focused and scoped to one topic. We evaluated in our backlogs that small stories were defined with fewer words, often with abbreviations and special words. In addition, high-quality small user stories are also often distinct in the sense of ignoring the user story format, which fosters a low-quality rated \textit{Format Complete} metric.

RQ2. Based on the results, we argue \textit{Format Complete} with the focus on more filled-in patterns and \textit{Word Sparse} with the orientation to an average word count representing the best metrics composition.

\section{Threats to Validity}
\label{section:limitations_future_work}

This section discusses internal and external threats to the validity of our expert study.

\subsection{Internal Validity}

A threat to internal validity might arise from the low quality of the expert ratings. To limit negative impacts, we chose experts familiar with user stories and their handling of agile software projects. Additionally, two pre-studies were conducted to improve our questionnaire and to find a reasonable amount of user story ratings per expert, to ensure the focus of the expert throughout the rating. We also integrated a tutorial at the start of our questionnaire to guide experts before they started the rating.

Finally, we chose guided sessions over a self-administered survey, to attain a high response rate and to support the experts concerning methodology questions, e.g., understanding the rating procedure. We used gathering experts' user story skills in years instead of categories such as ``Beginner'' and ``Advanced'', which reduce scale complexity and ensure a concise interpretation of the skill level by the experts. In addition, we verified outliers and multicollinearity to lower possible side effects on the regression analysis.

\subsection{External Validity}

We gathered similar results for USeR's lexical quality dimensions: \textit{Format Complete} and \textit{Word Sparse} regarding existing literature results \cite{Lucassen:2016im}, which can be better assessed and are better qualified to measure user story quality. Lexical quality dimensions are a common quality aspect in user stories, and we assume that the results of our study are generalizable to other user stories and projects.

Our study includes user stories from two real-world projects. To increase the generalizability of our results, we selected two different projects from divergent customer domains. In addition, we chose experts from different roles within the Scrum team, including quality assurance specialists, developers, and Product Owners. This acknowledges the existence of different perspectives on user story quality from team members concerned with writing, implementing, testing, and managing user stories.

We used a limited number of 50 user stories for each automotive and health evaluation, which could negatively influence the regression model parameter estimations. To mitigate possible negative side effects, we ran a power analysis beforehand to identify the appropriate number of user stories needed for a sound model evaluation. Finally, we reviewed the literature on expert studies to ensure the number of experts was sufficient for our study.

\section{Conclusions and Future Work}
\label{section:conclusions}

In this paper, we presented USeR, a web-based user story eReviewer that allows authors, such as Product Owners in agile software projects, to assess and optimize user story quality through a user interface for instant, consistent, and explainable user feedback, supporting fast and easy quality optimizations.

USeR's development process was covered with the (i) collection of potential quality metrics, (ii) refinement process to narrow down the initial raw metrics, (iii) derivation of algorithms, and finally, we provided (iv) USeR's web-based user interface, API, and environment.

We evaluated the effectiveness of USeR in an expert study with 100 user stories and four experts from two real-world agile software projects in the automotive and health sectors and used multiple linear regression for our analysis. Our results support USeR as a valid assistance tool for user story quality. The focus on more filled-in patterns and orientation to an average word count covers the best consistent metrics composition for USeR.

The concept of USeR shall lay the foundations for tools to support authors during the daily writing of user stories. We wish to inspire future research on intelligent assistants, for USeR, we plan features of (i) highlighting inline low-quality issues in user stories while (ii) providing recommendations to improve them, with (iii) support of Large Language Models (LLMs). In addition, we plan to provide an (iv) mobile and (v) plugin version (e.g., Jira \cite{Atlassian:2002ji}) of USeR to seamlessly integrate optimizations of user stories into the daily work of agile authors.

\section*{Acknowledgment}

We thank all the experts who participated in our study for their time spent rating our user stories and for providing feedback. Finally, we thank the two companies that charitably provided user stories for our research.

\bibliographystyle{IEEEtran}
\bibliography{IEEEabrv,arxiv_2025_user}

\begin{thebibliography}{10}
\providecommand{\url}[1]{#1}
\csname url@samestyle\endcsname
\providecommand{\newblock}{\relax}
\providecommand{\bibinfo}[2]{#2}
\providecommand{\BIBentrySTDinterwordspacing}{\spaceskip=0pt\relax}
\providecommand{\BIBentryALTinterwordstretchfactor}{4}
\providecommand{\BIBentryALTinterwordspacing}{\spaceskip=\fontdimen2\font plus
\BIBentryALTinterwordstretchfactor\fontdimen3\font minus
  \fontdimen4\font\relax}
\providecommand{\BIBforeignlanguage}[2]{{%
\expandafter\ifx\csname l@#1\endcsname\relax
\typeout{** WARNING: IEEEtran.bst: No hyphenation pattern has been}%
\typeout{** loaded for the language `#1'. Using the pattern for}%
\typeout{** the default language instead.}%
\else
\language=\csname l@#1\endcsname
\fi
#2}}
\providecommand{\BIBdecl}{\relax}
\BIBdecl

\bibitem{Agilealliance:2022us}
\BIBentryALTinterwordspacing
Agilealliance.org. (2022) User story template. Agile Alliance. [Online].
  Available: \url{https://www.agilealliance.org/glossary/user-story-template}
\BIBentrySTDinterwordspacing

\bibitem{Schwaber:2020th}
\BIBentryALTinterwordspacing
K.~Schwaber and J.~Sutherland. (2020) The scrum guide. [Online]. Available:
  \url{https://www.scrumguides.org/docs/scrumguide/v2020/2020-Scrum-Guide-US.pdf}
\BIBentrySTDinterwordspacing

\bibitem{Jeffries:2001es}
\BIBentryALTinterwordspacing
R.~Jeffries. (2001) Essential {XP}: Card, conversation, confirmation. [Online].
  Available:
  \url{https://www.ronjeffries.com/xprog/articles/expcard\\conversationconfirmation}
\BIBentrySTDinterwordspacing

\bibitem{Wake:2003in}
\BIBentryALTinterwordspacing
B.~Wake. (2003) {INVEST} in good stories, and {SMART} tasks. [Online].
  Available:
  \url{https://www.xp123.com/articles/invest-in-good-stories-and-smart-tasks}
\BIBentrySTDinterwordspacing

\bibitem{Cohn:2004us}
M.~Cohn, \emph{User Stories Applied: For Agile Software Development}, ser.
  Addison-Wesley Professional.\hskip 1em plus 0.5em minus 0.4em\relax Boston,
  MA, USA: Addison-Wesley Professional, 03 2004.

\bibitem{Wake:2012as}
\BIBentryALTinterwordspacing
B.~Wake. (2012) {``As a Developer{\dots}''} is not a user story. [Online].
  Available:
  \url{http://www.industriallogic.com/blog/as-a-developer-is-not-a-user-story}
\BIBentrySTDinterwordspacing

\bibitem{Pichler:20135c}
\BIBentryALTinterwordspacing
R.~Pichler. (2013) 5 common user story mistakes. [Online]. Available:
  \url{https://www.romanpichler.com/blog/5-common-user-story-mistakes}
\BIBentrySTDinterwordspacing

\bibitem{Converse:1993wq}
J.~A. Cannon-Bowers, E.~Salas, and S.~A. Converse, ``Shared mental models in
  expert team decision making,'' \emph{Individual and Group Decision Making:
  Current Issues}, vol. 221, pp. 221--246, 1993.

\bibitem{Sommerville:2005an}
\BIBentryALTinterwordspacing
I.~Sommerville and J.~Ransom, ``An empirical study of industrial requirements
  engineering process assessment and improvement,'' \emph{{ACM} Trans. Softw.
  Eng. Methodol.}, vol.~14, no.~1, pp. 85--117, 2005. [Online]. Available:
  \url{https://doi.org/10.1145/1044834.1044837}
\BIBentrySTDinterwordspacing

\bibitem{Lucassen:2016th}
\BIBentryALTinterwordspacing
G.~Lucassen, F.~Dalpiaz, J.~M. E.~M. van~der Werf, and S.~Brinkkemper, ``The
  use and effectiveness of user stories in practice,'' in \emph{Requirements
  Engineering: Foundation for Software Quality - 22nd International Working
  Conference, {REFSQ} 2016, Gothenburg, Sweden, March 14-17, 2016,
  Proceedings}, ser. Lecture Notes in Computer Science, M.~Daneva and
  O.~Pastor, Eds., vol. 9619.\hskip 1em plus 0.5em minus 0.4em\relax Springer,
  2016, pp. 205--222. [Online]. Available:
  \url{https://doi.org/10.1007/978-3-319-30282-9\_14}
\BIBentrySTDinterwordspacing

\bibitem{Femmer:2019re}
\BIBentryALTinterwordspacing
H.~Femmer and A.~Vogelsang, ``Requirements quality is quality in use,''
  \emph{{IEEE} Softw.}, vol.~36, no.~3, pp. 83--91, 2019. [Online]. Available:
  \url{https://doi.org/10.1109/MS.2018.110161823}
\BIBentrySTDinterwordspacing

\bibitem{Lucassen:2015fo}
\BIBentryALTinterwordspacing
G.~Lucassen, F.~Dalpiaz, J.~M. E.~M. van~der Werf, and S.~Brinkkemper,
  ``Forging high-quality user stories: Towards a discipline for agile
  requirements,'' in \emph{23rd {IEEE} International Requirements Engineering
  Conference, {RE} 2015, Ottawa, ON, Canada, August 24-28, 2015}, D.~Zowghi,
  V.~Gervasi, and D.~Amyot, Eds.\hskip 1em plus 0.5em minus 0.4em\relax {IEEE}
  Computer Society, 2015, pp. 126--135. [Online]. Available:
  \url{https://doi.org/10.1109/RE.2015.7320415}
\BIBentrySTDinterwordspacing

\bibitem{Lucassen:2016im}
\BIBentryALTinterwordspacing
------, ``Improving agile requirements: The quality user story framework and
  tool,'' \emph{Requir. Eng.}, vol.~21, no.~3, pp. 383--403, 2016. [Online].
  Available: \url{https://doi.org/10.1007/s00766-016-0250-x}
\BIBentrySTDinterwordspacing

\bibitem{Lucassen:2017im}
\BIBentryALTinterwordspacing
------, ``Improving user story practice with the grimm method: {A} multiple
  case study in the software industry,'' in \emph{Requirements Engineering:
  Foundation for Software Quality - 23rd International Working Conference,
  {REFSQ} 2017, Essen, Germany, February 27 - March 2, 2017, Proceedings}, ser.
  Lecture Notes in Computer Science, P.~Gr{\"{u}}nbacher and A.~Perini, Eds.,
  vol. 10153.\hskip 1em plus 0.5em minus 0.4em\relax Springer, 2017, pp.
  235--252. [Online]. Available:
  \url{https://doi.org/10.1007/978-3-319-54045-0\_18}
\BIBentrySTDinterwordspacing

\bibitem{Hayes:2015me}
\BIBentryALTinterwordspacing
J.~H. Hayes, W.~Li, T.~Yu, X.~Han, M.~Hays, and C.~Woodson, ``Measuring
  requirement quality to predict testability,'' in \emph{2015 {IEEE} Second
  International Workshop on Artificial Intelligence for Requirements
  Engineering, {AIRE} 2015, Ottawa, ON, Canada, August 24, 2015}.\hskip 1em
  plus 0.5em minus 0.4em\relax {IEEE} Computer Society, 2015, pp. 1--8.
  [Online]. Available: \url{https://doi.org/10.1109/AIRE.2015.7337622}
\BIBentrySTDinterwordspacing

\bibitem{Hallmann:2020id}
D.~Hallmann, ``{I Don't Understand!''}: Toward a model to evaluate the role of
  user story quality,'' in \emph{Proceedings of the 21st International
  Conference on Agile Software Development, {XP} 2020}, ser. Lecture Notes in
  Business Information Processing, vol. 383.\hskip 1em plus 0.5em minus
  0.4em\relax Copenhagen, Denmark: Springer, June 8-12 2020, pp. 103--112.

\bibitem{Lai:2017au}
S.-T. Lai, ``A user story quality measurement model for reducing agile software
  development risk,'' \emph{Int. J. Softw. Eng. \& Appl.}, vol.~8, no.~2, pp.
  75--86, 2017.

\bibitem{Lindland:1994un}
\BIBentryALTinterwordspacing
O.~I. Lindland, G.~Sindre, and A.~S{\o}lvberg, ``Understanding quality in
  conceptual modeling,'' \emph{{IEEE} Softw.}, vol.~11, no.~2, pp. 42--49,
  1994. [Online]. Available: \url{https://doi.org/10.1109/52.268955}
\BIBentrySTDinterwordspacing

\bibitem{Reimers:2019se}
\BIBentryALTinterwordspacing
N.~Reimers and I.~Gurevych, ``{Sentence-BERT}: Sentence embeddings using
  siamese bert-networks,'' in \emph{Proceedings of the 2019 Conference on
  Empirical Methods in Natural Language Processing and the 9th International
  Joint Conference on Natural Language Processing, {EMNLP-IJCNLP} 2019, Hong
  Kong, China, November 3-7, 2019}, K.~Inui, J.~Jiang, V.~Ng, and X.~Wan,
  Eds.\hskip 1em plus 0.5em minus 0.4em\relax Association for Computational
  Linguistics, 2019, pp. 3980--3990. [Online]. Available:
  \url{https://doi.org/10.18653/v1/D19-1410}
\BIBentrySTDinterwordspacing

\bibitem{Aaron:2024ge}
\BIBentryALTinterwordspacing
A.~Chibb. {German\_Semantic\_STS\_V2:\ Sentence-Transformer Model - Hugging
  Face}. [Online]. Available:
  \url{https://huggingface.co/aari1995/\\German\_Semantic\_STS\_V2}
\BIBentrySTDinterwordspacing

\bibitem{Lent:2004ae}
M.~van Lent, W.~Fisher, and M.~Mancuso, ``An explainable artificial
  intelligence system for small-unit tactical behavior,'' in \emph{Proceedings
  of the Nineteenth National Conference on Artificial Intelligence, Sixteenth
  Conference on Innovative Applications of Artificial Intelligence, July 25-29,
  2004, San Jose, California, {USA}}, D.~L. McGuinness and G.~Ferguson,
  Eds.\hskip 1em plus 0.5em minus 0.4em\relax {AAAI} Press / The {MIT} Press,
  2004, pp. 900--907.

\bibitem{Tickle:1998th}
A.~B. Tickle, R.~Andrews, M.~Golea, and J.~Diederich, ``The truth will come to
  light: Directions and challenges in extracting the knowledge embedded within
  trained artificial neural networks,'' \emph{{IEEE} Trans. Neural Netw.},
  vol.~9, no.~6, pp. 1057--1068, 1998.

\bibitem{Edwards:2017sl}
L.~Edwards and M.~Veale, ``Slave to the algorithm? why a'right to an
  explanation'is probably not the remedy you are looking for,'' \emph{Duke L.
  \& Tech. Rev.}, vol.~16, p.~18, 2017.

\bibitem{Miranda:2009si}
\BIBentryALTinterwordspacing
E.~Miranda, P.~Bourque, and A.~Abran, ``Sizing user stories using paired
  comparisons,'' \emph{Inf. Softw. Technol.}, vol.~51, no.~9, pp. 1327--1337,
  2009. [Online]. Available: \url{https://doi.org/10.1016/j.infsof.2009.04.003}
\BIBentrySTDinterwordspacing

\bibitem{Abrahamsson:2011pr}
\BIBentryALTinterwordspacing
P.~Abrahamsson, I.~Fronza, R.~Moser, J.~Vlasenko, and W.~Pedrycz, ``Predicting
  development effort from user stories,'' in \emph{Proceedings of the 5th
  International Symposium on Empirical Software Engineering and Measurement,
  {ESEM} 2011, Banff, AB, Canada, September 22-23, 2011}.\hskip 1em plus 0.5em
  minus 0.4em\relax {IEEE} Computer Society, 2011, pp. 400--403. [Online].
  Available: \url{https://doi.org/10.1109/ESEM.2011.58}
\BIBentrySTDinterwordspacing

\bibitem{Raith:2013id}
\BIBentryALTinterwordspacing
F.~Raith, I.~Richter, R.~Lindermeier, and G.~Klinker, ``Identification of
  inaccurate effort estimates in agile software development,'' in \emph{20th
  Asia-Pacific Software Engineering Conference, {APSEC} 2013, Ratchathewi,
  Bangkok, Thailand, December 2-5, 2013 - Volume 2}, P.~Muenchaisri and
  G.~Rothermel, Eds.\hskip 1em plus 0.5em minus 0.4em\relax {IEEE} Computer
  Society, 2013, pp. 67--72. [Online]. Available:
  \url{https://doi.org/10.1109/APSEC.2013.114}
\BIBentrySTDinterwordspacing

\bibitem{Popli:2014co}
R.~Popli and N.~Chauhan, ``Cost and effort estimation in agile software
  development,'' in \emph{Proceedings of the 2014 International Conference on
  Optimization, Reliabilty, and Information Technology, {ICROIT} 2014}.\hskip
  1em plus 0.5em minus 0.4em\relax Faridabad, India: {IEEE} Computer Society,
  February 6-8 2014, pp. 57--61.

\bibitem{Popli:2014es}
------, ``Estimation in agile environment using resistance factors,'' in
  \emph{Proceedings of the 2014 International Conference on Information Systems
  and Computer Networks, {ISCON} 2014}.\hskip 1em plus 0.5em minus 0.4em\relax
  Mathura, India: {IEEE} Computer Society, March 1-2 2014, pp. 60--65.

\bibitem{Stettina:2011ne}
\BIBentryALTinterwordspacing
C.~J. Stettina and W.~Heijstek, ``Necessary and neglected?: an empirical study
  of internal documentation in agile software development teams,'' in
  \emph{Proceedings of the 29th {ACM} international conference on Design of
  communication, Pisa, Italy, October 3-5, 2011}, A.~Protopsaltis, N.~Spyratos,
  C.~J. Costa, and C.~Meghini, Eds.\hskip 1em plus 0.5em minus 0.4em\relax
  {ACM}, 2011, pp. 159--166. [Online]. Available:
  \url{https://doi.org/10.1145/2038476.2038509}
\BIBentrySTDinterwordspacing

\bibitem{Stettina:2012do}
\BIBentryALTinterwordspacing
C.~J. Stettina, W.~Heijstek, and T.~E. F{\ae}gri, ``Documentation work in agile
  teams: The role of documentation formalism in achieving a sustainable
  practice,'' in \emph{2012 Agile Conference, {AGILE} 2012, Dallas, TX, USA,
  August 13-17, 2012}.\hskip 1em plus 0.5em minus 0.4em\relax {IEEE} Computer
  Society, 2012, pp. 31--40. [Online]. Available:
  \url{https://doi.org/10.1109/Agile.2012.7}
\BIBentrySTDinterwordspacing

\bibitem{Sharp:2009th}
\BIBentryALTinterwordspacing
H.~Sharp, H.~Robinson, and M.~Petre, ``The role of physical artefacts in agile
  software development: Two complementary perspectives,'' \emph{Interact.
  Comput.}, vol.~21, no. 1-2, pp. 108--116, 2009. [Online]. Available:
  \url{https://doi.org/10.1016/j.intcom.2008.10.006}
\BIBentrySTDinterwordspacing

\bibitem{Beck:2001ma}
\BIBentryALTinterwordspacing
K.~Beck, M.~Beedle, A.~V. Bennekum, A.~Cockburn, W.~Cunningham, M.~Fowler,
  J.~Highsmith, A.~Hunt, R.~Jeffries, J.~Kern, B.~Marick, R.~C. Martin,
  K.~Schwaber, J.~Sutherland, and D.~Thomas. (2001) {Manifesto for Agile
  Software Development}. [Online]. Available: \url{http://agilemanifesto.org}
\BIBentrySTDinterwordspacing

\bibitem{DeSouza:2005as}
\BIBentryALTinterwordspacing
S.~C.~B. de~Souza, N.~Anquetil, and K.~M. de~Oliveira, ``A study of the
  documentation essential to software maintenance,'' in \emph{Proceedings of
  the 23rd Annual International Conference on Design of Communication:
  documenting {\&} Designing for Pervasive Information, {SIGDOC} 2005,
  Coventry, UK, September 21-23, 2005}, S.~R. Tilley and R.~M. Newman,
  Eds.\hskip 1em plus 0.5em minus 0.4em\relax {ACM}, 2005, pp. 68--75.
  [Online]. Available: \url{https://doi.org/10.1145/1085313.1085331}
\BIBentrySTDinterwordspacing

\bibitem{Flesch:1948an}
R.~Flesch, ``A new readability yardstick.'' \emph{J. Appl. Psychol.}, vol.~32,
  no.~3, p. 221, 1948.

\bibitem{Stettina:2013th}
C.~J. Stettina and E.~Kroon, ``Is there an agile handover? an empirical study
  of documentation and project handover practices across agile software
  teams,'' in \emph{2013 International Conference on Engineering, Technology
  and Innovation (ICE) \& IEEE International Technology Management Conference},
  ser. International Conference on Engineering, Technology and Innovation \&
  IEEE International Technology Management Conference, 2013, pp. 1 -- 12.

\bibitem{Boehm:1984ve}
\BIBentryALTinterwordspacing
B.~W. Boehm, ``Verifying and validating software requirements and design
  specifications,'' \emph{{IEEE} Softw.}, vol.~1, no.~1, pp. 75--88, 1984.
  [Online]. Available: \url{https://doi.org/10.1109/MS.1984.233702}
\BIBentrySTDinterwordspacing

\bibitem{Davis:1993sr}
A.~M. Davis, \emph{Software Requirements - objects, functions, and states},
  ser. Prentice Hall international editions.\hskip 1em plus 0.5em minus
  0.4em\relax Prentice Hall, 1993.

\bibitem{Iso:2018is}
ISO, ``{ISO}/{IEC}/{IEEE} {International Standard - Systems and software
  engineering -- Life cycle processes -- Requirements engineering},''
  \emph{ISO/IEC/IEEE 29148:2018(E)}, pp. 1--104, 2018.

\bibitem{Malone:1994th}
\BIBentryALTinterwordspacing
T.~W. Malone and K.~Crowston, ``The interdisciplinary study of coordination,''
  \emph{{ACM} Comput. Surv.}, vol.~26, no.~1, pp. 87--119, 1994. [Online].
  Available: \url{https://doi.org/10.1145/174666.174668}
\BIBentrySTDinterwordspacing

\bibitem{Atlassian:2002ji}
\BIBentryALTinterwordspacing
(2002) {JIRA Software - Issue \& Project Tracking for Software Teams |
  Atlassian}. [Online]. Available:
  \url{https://www.atlassian.com/software/jira}
\BIBentrySTDinterwordspacing

\bibitem{Langacker:1987fo}
R.~W. Langacker, \emph{Foundations of cognitive grammar: Volume I: Theoretical
  prerequisites}.\hskip 1em plus 0.5em minus 0.4em\relax Stanford university
  press, 1987, vol.~1.

\bibitem{ZSL:2023gr}
\BIBentryALTinterwordspacing
{Baden-W{\"u}rttenberg, Center for School Quality and Teacher Training. (Author
  translated to English)}. (2024) {Basic vocabulary German elementary school.
  (Title translated to English)}. [Online]. Available:
  \url{https://zsl-bw.de/,Lde/Startseite/allgemeine-bildung/grundwortschatz-deutsch-gs}
\BIBentrySTDinterwordspacing

\bibitem{NodeJS:2024no}
\BIBentryALTinterwordspacing
{OpenJS Foundation}. (2024) {Node.js}. [Online]. Available:
  \url{https://nodejs.org}
\BIBentrySTDinterwordspacing

\bibitem{React:2024no}
\BIBentryALTinterwordspacing
{MetaOpenSource}. (2024) {React - The library for web and native user
  interfaces}. [Online]. Available: \url{https://react.dev}
\BIBentrySTDinterwordspacing

\bibitem{Python:2024no}
\BIBentryALTinterwordspacing
{Python Software Foundation}. (2024) {Python}. [Online]. Available:
  \url{https://python.org}
\BIBentrySTDinterwordspacing

\bibitem{SpaCy:2024sp}
\BIBentryALTinterwordspacing
{ExplosionAI GmbH}. {SpaCy - Industrial-Strength Natural Language Processing}.
  [Online]. Available: \url{https://spacy.io}
\BIBentrySTDinterwordspacing

\bibitem{scikit-learn:2011sc}
F.~Pedregosa, G.~Varoquaux, A.~Gramfort, V.~Michel, B.~Thirion, O.~Grisel,
  M.~Blondel, P.~Prettenhofer, R.~Weiss, V.~Dubourg, J.~Vanderplas, A.~Passos,
  D.~Cournapeau, M.~Brucher, M.~Perrot, and {\'E}.~Duchesnay, ``{Scikit-learn:
  Machine Learning in {P}ython},'' \emph{J. Mach. Learn. Res.}, vol.~12, pp.
  2825--2830, 2011.

\bibitem{Grootendorst:2022be}
M.~Grootendorst, ``{BERTopic}: Neural topic modeling with a class-based tf-idf
  procedure,'' \emph{arXiv preprint arXiv:2203.05794}, 2022.

\bibitem{Docker:2024no}
\BIBentryALTinterwordspacing
{Docker Inc.} (2024) {Docker}. [Online]. Available: \url{https://docker.com}
\BIBentrySTDinterwordspacing

\bibitem{Arora:2016ex}
\BIBentryALTinterwordspacing
C.~Arora, M.~Sabetzadeh, L.~C. Briand, and F.~Zimmer, ``Extracting domain
  models from natural-language requirements: approach and industrial
  evaluation,'' in \emph{Proceedings of the {ACM/IEEE} 19th International
  Conference on Model Driven Engineering Languages and Systems, Saint-Malo,
  France, October 2-7, 2016}, B.~Baudry and B.~Combemale, Eds.\hskip 1em plus
  0.5em minus 0.4em\relax {ACM}, 2016, pp. 250--260. [Online]. Available:
  \url{http://dl.acm.org/citation.cfm?id=2976769}
\BIBentrySTDinterwordspacing

\bibitem{Israel:2008in}
J.~H. Israel, C.~Z{\"{o}}llner, M.~Mateescu, R.~Korkot, G.~Bittersmann, P.~T.
  Fischer, J.~Neumann, and R.~Stark, ``Investigating user requirements and
  usability of immersive three-dimensional sketching for early conceptual
  design - results from expert discussions and user studies,'' in
  \emph{Proceedings of the 5th Eurographics Workshop on Sketch-Based Interfaces
  and Modeling, {SBIM} 2008}.\hskip 1em plus 0.5em minus 0.4em\relax Annecy,
  France: Eurographics Association, June 11-13 2008, pp. 127--134.

\bibitem{Cohen:1988st}
J.~Cohen, \emph{Statistical Power Analysis for the Behavioral Sciences}.\hskip
  1em plus 0.5em minus 0.4em\relax Taylor \& Francis, 1988.

\bibitem{Gpower:2009st}
F.~Faul, E.~Erdfelder, A.~Buchner, and A.-G. Lang, ``Statistical power analyses
  using {G* Power 3.1}: Tests for correlation and regression analyses,''
  \emph{Behav. Res. Methods}, vol.~41, no.~4, pp. 1149--1160, 2009.

\bibitem{Casella:2024st}
G.~Casella and R.~Berger, \emph{Statistical inference}.\hskip 1em plus 0.5em
  minus 0.4em\relax CRC press, 2024.

\bibitem{Oates:2006re}
B.~J. Oates, \emph{Researching Information Systems and Computing}.\hskip 1em
  plus 0.5em minus 0.4em\relax Los Angeles, California, USA: SAGE Publications,
  Inc., 2006.

\bibitem{Creswell:2017re}
J.~W. Creswell and J.~D. Creswell, \emph{Research design: Qualitative,
  quantitative, and mixed methods approaches}.\hskip 1em plus 0.5em minus
  0.4em\relax Sage publications, 2017.

\bibitem{Likert:1932at}
R.~Likert, ``A technique for the measurement of attitudes,'' \emph{Arch.
  Psychol.}, vol.~22, pp. 5--55, 1932.

\bibitem{Limesurvey:2022li}
\BIBentryALTinterwordspacing
{LimeSurvey GmbH}. (2022) {LimeSurvey}. [Online]. Available:
  \url{https://www.limesurvey.org}
\BIBentrySTDinterwordspacing

\bibitem{Zoom:2022zo}
\BIBentryALTinterwordspacing
{Zoom Video Communications, Inc.} (2022) {Zoom}. [Online]. Available:
  \url{https://www.zoom.us}
\BIBentrySTDinterwordspacing

\bibitem{Myrtveit:1999ac}
\BIBentryALTinterwordspacing
I.~Myrtveit and E.~Stensrud, ``A controlled experiment to assess the benefits
  of estimating with analogy and regression models,'' \emph{{IEEE} Trans.
  Softw. Eng.}, vol.~25, no.~4, pp. 510--525, 1999. [Online]. Available:
  \url{https://doi.org/10.1109/32.799947}
\BIBentrySTDinterwordspacing

\bibitem{Gisev:2013in}
\BIBentryALTinterwordspacing
N.~Gisev, J.~S. Bell, and T.~F. Chen, ``Interrater agreement and interrater
  reliability: Key concepts, approaches, and applications,'' \emph{Research in
  Social and Administrative Pharmacy}, vol.~9, no.~3, pp. 330--338, 2013.
  [Online]. Available:
  \url{https://www.sciencedirect.com/science/article/pii/S1551741112000642}
\BIBentrySTDinterwordspacing

\bibitem{SmartPLS:2024no}
\BIBentryALTinterwordspacing
{SmartPLS GmbH}. (2024) {SmartPLS - Next Generation Path Modeling}. [Online].
  Available: \url{https://www.smartpls.com}
\BIBentrySTDinterwordspacing

\bibitem{Barnett:1994ou}
V.~Barnett, T.~Lewis \emph{et~al.}, \emph{Outliers in statistical data}.\hskip
  1em plus 0.5em minus 0.4em\relax Wiley New York, 1994, vol.~3, no.~1.

\bibitem{Hoaglin:2003st}
D.~C. Hoaglin, ``John w. tukey and data analysis,'' \emph{Statistical Science},
  pp. 311--318, 2003.

\bibitem{Fox:2018an}
J.~Fox and S.~Weisberg, \emph{An R companion to applied regression}.\hskip 1em
  plus 0.5em minus 0.4em\relax Sage publications, 2018.

\end{thebibliography}

\end{document}